\documentclass[conference]{IEEEtran}

\IEEEoverridecommandlockouts
\usepackage{subcaption}
\usepackage{caption}
\usepackage{verbatim}
\usepackage{graphicx}
\usepackage{cite}
\usepackage{url}
\usepackage[cmex10]{amsmath}
\usepackage{amssymb}
\usepackage{algorithm}
\usepackage{algpseudocode}
\usepackage{cases}
\usepackage{color}
\usepackage{cite}
\usepackage{epstopdf}
\usepackage{calc}
\usepackage{array}
\usepackage{algpseudocode}
\usepackage{xparse}
\let\oldState\State
\RenewDocumentCommand{\State}{o}{
	\IfValueTF{#1}{\makeatletter\setcounter{ALG@line}{#1}\addtocounter{ALG@line}{-1}\makeatother}{}%
	\oldState\ignorespaces%
}%

\DeclareGraphicsExtensions{.pdf,.eps,.png,.jpg,.mps}

\begin{document}
	\title{\huge Unsupervised Frequency Clustering Algorithm for Null Space Estimation in Wideband Spectrum Sharing Networks}
	\author{\IEEEauthorblockN{Shailesh Chaudhari and Danijela Cabric}
		\IEEEauthorblockA{Department of Electrical Engineering,
			University of California, Los Angeles\\
			Email: schaudhari@ucla.edu, danijela@ee.ucla.edu}
		\vspace{-10mm}
	}
	\maketitle

\begin{abstract}
	In spectrum sharing networks, a base station (BS) needs to mitigate the interference to users associated with other coexisting network in the same band. The BS can achieve this by transmitting its downlink signal in the null space of channels to such users. However, under a wideband scenario, the BS needs to estimate null space matrices using the received signal from such non-cooperative users in each frequency bin where the users are active. To reduce the computational complexity of this operation, we propose a frequency clustering algorithm that exploits the channel correlations among adjacent frequency bins. The proposed algorithm forms clusters of frequency bins with correlated channel vectors without prior knowledge of the channels and obtains a single null space matrix for each cluster. We show that the number of matrices and the number of eigenvalue decompositions required to obtain the null space significantly reduce using the proposed clustering algorithm.
	
\end{abstract}
\IEEEpeerreviewmaketitle

\section{Introduction}
\label{sec:intro}
Spectrum sharing networks allow multiple networks to coexist in the same frequency band in order to improve spectrum efficiency. Coexistence of LTE-U and WiFi networks is one such example of spectrum sharing networks \cite{chen2016, yun2015}. In such networks, a base station (BS) equipped with multiple antennas can mitigate the interference to users associated with the other coexisting network by transmitting its downlink signal in the null space of channels to users in the other network \cite{yun2015,tsinos2013a,perlaza2010}. We refer to users associated with the other network as Outside Users (OUs)\footnote{In coexisting LTE-U and WiFi networks,  the users  associated with coexisting WiFi network are OUs with respect to the LTE BS.}. The null space matrix can be estimated at the BS without any cooperation by observing the received signal from OUs and computing the eigenvectors corresponding to noise eigenvalues \cite{gao2010a, tsinos2013a}. In a wideband channel, the BS channelizes the received signal using Fast Fourier Transform (FFT) and then computes null space matrices in each FFT bin with active OU signal \cite{ghamariadian2014,kouassi2013}. This brute-force approach becomes computationally expensive as the number of null space matrices and the number of eigenvalue decompositions (EVDs)  are equal to the number of frequency bins where OUs are active.

In this paper, we propose a low complexity algorithm to estimate the null space for wideband channels by clustering the frequency bins with correlated channels. The correlation among the channels  arises due to the fact that the number of taps in wideband multipath channels are usually less than the number of FFT points \cite{choi2005, sanchez-garcia2009}. We propose a new test statistic to form clusters of correlated bins. A single null space matrix is then computed for each cluster. The proposed unsupervised   algorithm does not require any prior knowledge of the channels or training signal, and significantly reduces the computational complexity of obtaining the null space of wideband channels.

The paper is organized as follows. The system model and problem formulation are provided in Section \ref{sec:system_model}. The test statistic for clustering as well as the clustering algorithm are presented in Section \ref{sec:proposed_method}. Section \ref{sec:simulation} provides simulation results under different channel models. Finally, concluding remarks are provided in Section \ref{sec:conclusion}.

\textit{Notations}:
Vectors are denoted by bold, lower-case letters, e.g., $\mathbf{h}$. Matrices are denoted by bold, upper case letters, e.g., $\mathbf{R}$. Hermitian transpose is denoted by $(.)^H$. The norm of a vector $\mathbf{h}$ is denoted by $||\mathbf{h}||$. The $i$-th element in a set $\mathcal{S}$ is denoted by $\mathcal{S}(i)$, while $|\mathcal{S}|$ denotes the cardinality of the set. Finally, a set of integers from $a$ to $b$ is denoted by $[a,b]$.

\section{System Model}
\label{sec:system_model}
\begin{figure}
	\centering
	\includegraphics[width= \columnwidth]{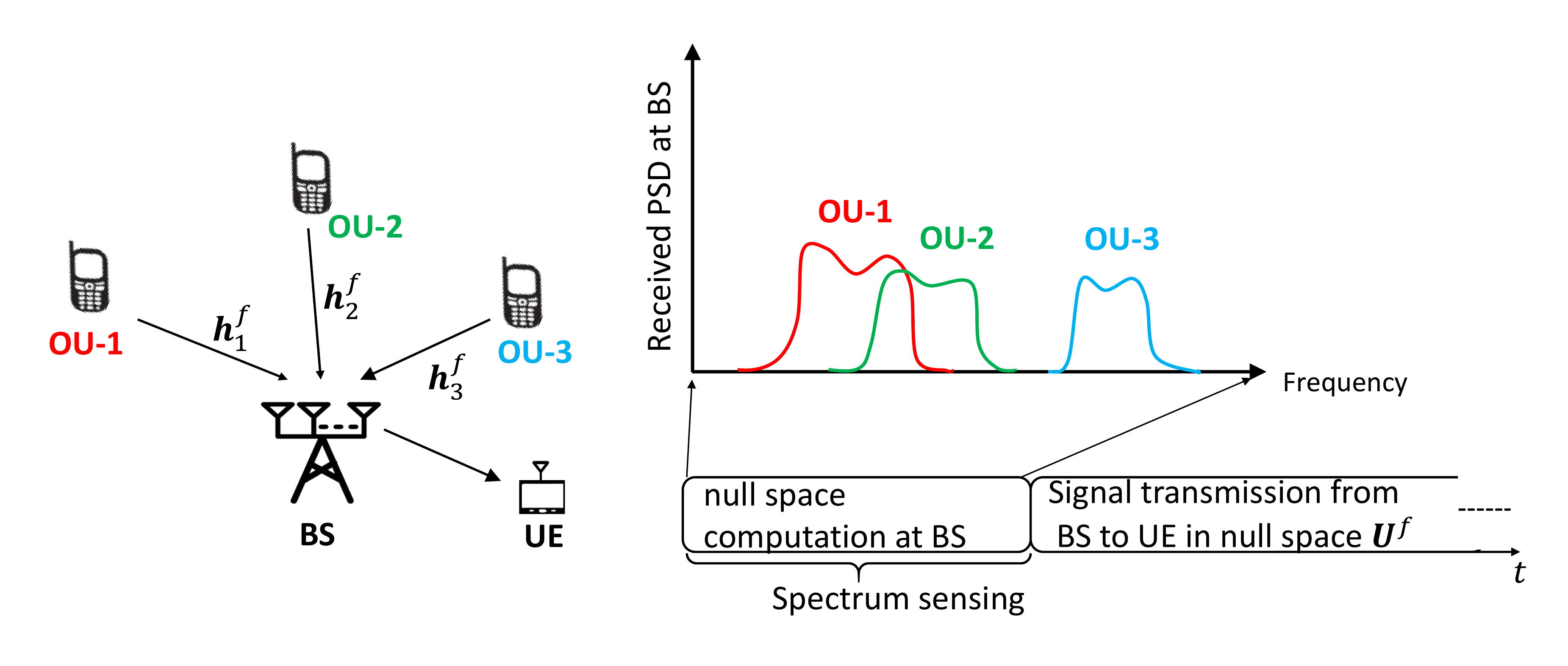}
	\caption{{\footnotesize System model: BS senses wideband signals received from $L=3$ OUs using $F$-point FFT.}}
	\label{fig:system_model}
	\vspace{-5mm}
\end{figure}

Consider that a BS, equipped with $M$ antennas, receives signals in a wideband spectrum from $L$ OUs at sampling rate $1/T_s$, as shown in Fig. \ref{fig:system_model}.  The BS channelizes the received signal by taking $F$-point FFT on each antenna. Let us denote the set of frequency bins with active OUs by $\mathcal{F}_a \subset [1,F]$. Then, the $M\times 1$ received signal vector in frequency bin $f\in \mathcal{F}_a$ is given by
\begin{align}
\mathbf{r}^f(n) = \sum_{l=1}^{L_f}\sqrt{p^f_l} x^f_l(n) \mathbf{h}^f_l + \mathbf{w}^f(n), f\in \mathcal{F}_a,
\end{align}
where $L^f \leq L$ is the number of OUs active in frequency bin $f$, $p^f_l$ and $x^f_l(t)$ are the transmitted power and symbol by OU-$l$, respectively. Further, $\mathbf{h}^f_l \in \mathbb{C}^{M\times 1}$ is channel vector between OU-$l$ and the BS\footnote{The $m$-th element of the  vector $\mathbf{h}^f_l$, say $h^f_{l,m}$, denotes the channel coefficient between OU-$l$ and the $m$-th antenna element in frequency bin $f$. If the time-domain $P$-tap multipath channel is $\mathbf{h}^t_l=[h^t_{l}(0),\cdots,h^t_{l}(P-1)]$, then the  $h^f_{l,m}$ is the $f$-th element in the $F$-point FFT of $\mathbf{h}^t_l$.}, and $\mathbf{w}^f(n)\sim CN(0,\sigma^2_{w} \mathbf{I})$ is the additive white Gaussian noise vector. We assume that the symbols are unit power, i.e., $\mathbb{E}[|x^f_l(t)|^2]=1$ and the channel $\mathbf{h}^f_l$ remains constant for $n=0,2,\cdots, T-1$. The covariance matrix of the vector $\mathbf{r}^f(n)$ is
{\small\begin{align}
\mathbf{R}^f = \mathbb{E}[\mathbf{r}^f(n) \mathbf{r}^f(n)^H] =  \sum_{l=1}^{L_f}p^f_l \mathbf{h}^f_l (\mathbf{h}^f_l)^H + \mathbf{R}^f_w, f\in \mathcal{F}_a,
\label{eq:cov}
\end{align}}
where $\mathbf{R}^f_w = \sigma^2_{w}\mathbf{I}$ is the noise covariance matrix. 

The channels between OUs and the BS are assumed to be reciprocal, which holds for time-division duplex systems. Therefore, in order to cancel the interference to OUs, the BS needs to transmit its signal in the orthogonal null space  $\mathbf{U}^f$ such that $||(\mathbf{U}^f)^H \mathbf{h}^f_l||=0$. The null space computation problem in the wideband can then be stated as follows:
\begin{align}
\nonumber (\mathbf{P1})~\text{Find}~~ &\mathbf{U}^f  
\\\text{Subject to}~~ &||(\mathbf{U}^f)^H \mathbf{h}^f_l|| = 0,~~ (\mathbf{U}^f)^H\mathbf{U}^f = \mathbf{I},
\\\nonumber &\mathbf{U}^f \in \mathbb{C}^{M \times (M-L_f)}, f \in \mathcal{F}_a, l=1,2,...,L_f.
\end{align}
Since the channel vectors $ \mathbf{h}^f_l$ are unknown at the BS, the null space matrices are obtained using the received covariance matrix by solving the following equivalent trace-difference minimization problem:
\begin{align}
\nonumber (\mathbf{P2})~\text{Min}_{\mathbf{U}^f}~~ &\text{Tr}((\mathbf{U}^f)^H (\mathbf{R}^f - \mathbf{R}^f_w)\mathbf{U}^f)
\\\text{Subject to}~~& (\mathbf{U}^f)^H\mathbf{U}^f = \mathbf{I},
\\\nonumber &\mathbf{U}^f \in \mathbb{C}^{M \times (M-L_f)},  f \in \mathcal{F}_a, l=1,2,...,L_f.
\end{align}
\noindent The solution of both problems $\mathbf{P1}$ and $\mathbf{P2}$ can be written as $\mathbf{U}^{f*} = \mathbf{V}^f_{L_f+1:M}$, where $\mathbf{V}^f \in \mathbb{C}^{M\times M}$ is a matrix containing eigenvectors of $\mathbf{R}^f - \mathbf{R}^f_w$ and $\mathbf{V}^f_{L_f+1:M}\in \mathbb{C}^{M\times M-L_f}$ is a matrix containing $L_f+1$ to $M$-th column vectors of $\mathbf{V}^f$, where the $m$-th column of $\mathbf{V}^f$ corresponds to $m$-th largest eigenvalue of  $(\mathbf{R}^f - \mathbf{R}^f_w)$. Note that the matrix $(\mathbf{R}^f - \mathbf{R}^f_w)$ has $M-L_f$ zero eigenvalues. Therefore, the optimum value of the objective function in $\mathbf{P2}$ is zero when $\mathbf{U}^f = \mathbf{U}^{f*}$. \footnote{Due to channel reciprocity, the optimal solution of $\mathbf{P2}$ ensures that $(\mathbf{U}^{f*})^H \mathbf{h}^f_l = 0$.}

Assuming that noise covariance matrix $\mathbf{R}^f_w$ is known, 
the brute-force method of obtaining the matrix $\mathbf{U}^f$ is to treat each frequency bin separately and compute $\mathbf{U}^f$ for each frequency bin using the received covariance matrix. This brute-force approach is an extension of the methods proposed in \cite{gao2010a,tsinos2013a} to a wideband scenario. This method first estimates the non-asymptotic signal covariance matrix using $T$ measurements:
\begin{align}
\mathbf{\hat{R}}^f = \frac{1}{T} \sum_{n=0}^{T-1}\mathbf{r}^f(n) (\mathbf{r}^f(n))^H
\label{eq:covariance_est}
\end{align}
Then, the null space matrix $\mathbf{U}^f$ is estimated from the EVD of $\mathbf{\hat{R}}^f - \mathbf{R}^f_w$. Note that in order to compute $\mathbf{U}^f$, the number of active OU signals $L_f$ is estimated from the eigenvalues of $\mathbf{\hat{R}}^f - \mathbf{R}^f_w$ using Wax-Kailath maximum description length estimator \cite{nadakuditi2008, wax1985}:
\begin{align}
	\hat{L}_{f} = \arg\min_{l \in [1,M]} (l-M)T \log\left( \frac{g(l)}{a(l)}\right) + 0.5 l(2M-l) \log(T),
\end{align}
where $g(l) = \prod_{j=l+1}^{M} d_j^{1/(M-l)}$ is the geometric mean of $M-l$ smallest eigenvalues of $(\mathbf{\hat{R}}^{f}-\mathbf{{R}}^{f}_w$), $a(l) = \frac{1}{M-l} \sum_{j=l+1}^{M}d_j$ is their arithmetic mean and $d_j$ is the $j$-th largest eigenvalue of $\mathbf{\hat{R}}^{f}-\mathbf{R}^{f}_w$. The number of matrices $\mathbf{U}^f$ required to be computed and the number of EVD computations required to obtain the solution of $\mathbf{P2}$ using brute-force wideband method is equal to number of active bins with active OUs: $|\mathcal{F}_a|$.

\section{Proposed method}
\label{sec:proposed_method}
\begin{figure}
	\centering
	\includegraphics[width=0.8\columnwidth]{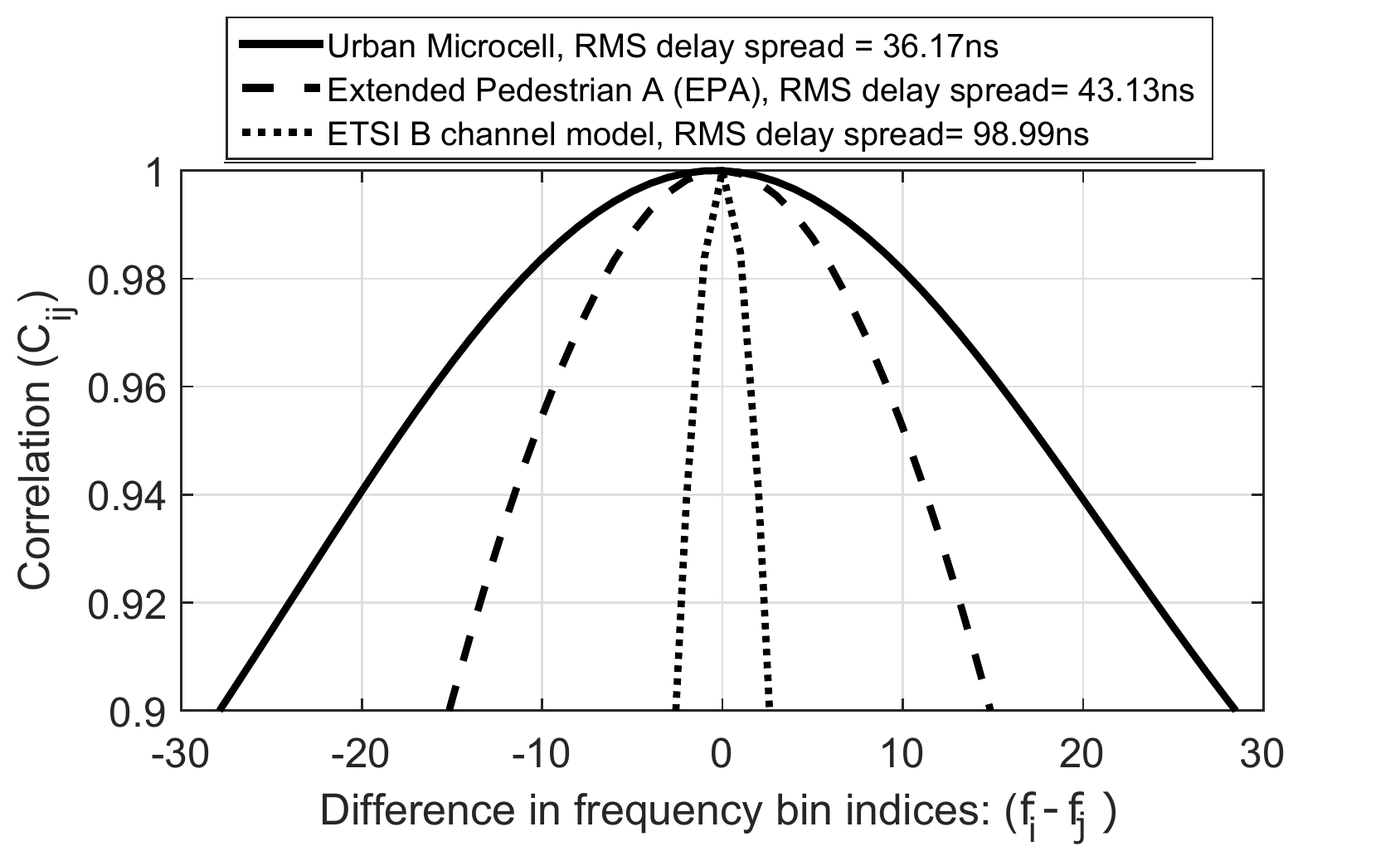}
	\caption{{\footnotesize Correlation $C_{ij}$ vs frequency bin spacing for three channel models with RMS delay spreads 36.17ns, 43.13ns, and 98.99ns. $1/T_s=20$MHz, $F=512$, $M=8$.}}
	\label{fig:channel_correlation}
	\vspace{-5mm}
\end{figure}

We exploit the correlation among the channel vectors $\mathbf{h}_l^f$ in adjacent frequency bins in order to reduce computational complexity of obtaining the null space matrices. The correlation among the channel vectors arises due to the fact that the number of taps in the multipath channels are usually less than the number of FFT points. The correlation coefficient in vectors on frequency bins $f_i$ and $f_j$ is defined as \cite{choi2005, sanchez-garcia2009}:
\begin{align}
C_{ij} = \frac{E\left[|(\mathbf{h}^{f_i}_l)^H  \mathbf{h}^{f_j}_l|^2\right]}{E\left[||\mathbf{h}^{f_i}_l||^2 \right]E \left[||\mathbf{h}^{f_j}_l||^2 \right]}
\end{align}
The correlation coefficients for different channel models is shown in Fig. \ref{fig:channel_correlation}. We can observe that the channels with lower RMS delay spreads have higher correlations $C_{ij}$ among adjacent frequency bins for same $T_s$ and $F$. We argue that the angle between $\mathbf{h}^{f_i}_l$ and $\mathbf{h}^{f_j}_l$ in one channel realization, given by $\theta_{ij}= \cos^{-1}\left(\frac{|(\mathbf{h}^{f_i}_l)^H  \mathbf{h}^{f_j}_l|}{||\mathbf{h}^{f_i}_l|| ||\mathbf{h}^{f_j}_l||}\right)$, approaches zero as $C_{ij}$ approaches 1. Therefore, the angle between the orthogonal null spaces $\mathbf{U}^{f_i}$ and $\mathbf{U}^{f_j}$ approaches zero as well. We can then describe the null space in bins $f_i$ and $f_j$ using one matrix $\mathbf{U}^{\mathcal{S}}$ without significant degradation in the relative quality of null defined as $Q^f_l=||(\mathbf{U}^{\mathcal{S}})^H \mathbf{h}^f_l||^2/ ||\mathbf{h}^f_l||^2, f\in \{f_i, f_j\}$. We propose to group such frequency bins with high correlation, equivalently small angle between their null spaces, in one cluster and find a common null space for the cluster. In order to decide whether to cluster frequency bin $f_j$ with $f_i$, we use the test statistic presented below. 

\subsection{Test statistic for clustering}
\label{section:test_statistic}
Let $p^{f}_r = \sum_{l=1}^{L_{f}}p^{f}_l||\mathbf{h}^{f}_l||^2 = \text{Tr}\left(\mathbf{{R}}^{f} -\mathbf{R}^{f}_w \right)$ be the power received on the antenna array in bin $f$. Further, let us define $D^{ij}_c= \text{Tr}\left((\mathbf{{U}}^{f_i})^H \left(\mathbf{{R}}^{f_j} -\mathbf{R}^{f_j}_w \right)\mathbf{{U}}^{f_i}\right) = (1-C_{ij})p^{f_j}_r$ as the component of $p^{f_j}_r$ in the span of $\mathbf{U}^{f_i}$. In other words, $D^{ij}_c$ is the square of length of projection of $\sum_l \sqrt{p^{f_j}_l}\mathbf{h}^{f_j}_l$ onto the space spanned by $\mathbf{U}^{f_i}$. We use this power component as the test statistic for clustering.

The value of $p^{f_j}_r$ is estimated as $\hat{p}^{f_j}_r =  \text{Tr}\left(\mathbf{\hat{R}}^{f} -\mathbf{R}^{f}_w \right)$. Similarly, the matrix $\mathbf{\hat{U}}^{f_i}$ is computed using EVD of the non-asymptotic signal covariance matrix $\mathbf{\hat{R}}^{f_i}$. Since $\hat{p}^{f_j}_r$ is the average received energy, it can be  modeled as a Gaussian random variable \cite{laghate2017a}: $\hat{p}^{f_j}_r =  \text{Tr}\left(\mathbf{\hat{R}}^{f} -\mathbf{R}^{f}_w \right) \sim \mathcal{N}(\mu_j, \sigma^2_j)$, where
{\small\begin{align}
	\nonumber \mu_{j} =\sum_{l=1}^{L_{f_j}}p^{f_j}_l||\mathbf{h}^{f_j}_l||^2, \text{and~~}
	\sigma^2_{j} = \frac{1}{T} \left[\left(\sum_{l=1}^{L_{f_j}}p^{f_j}_l||\mathbf{h}^{f_j}_l||^2 \right)^2 + \sigma^4_w \right].
	\end{align}}
 Therefore, the proposed test statistic is modeled as a Gaussian random variable:
\begin{align}
	\hat{D}^{ij}_c \sim \mathcal{N}\left((1-C_{ij}) \mu_j, (1-C_{ij})^2 \sigma^2_j\right).
\end{align}

Let us consider that we would like to cluster the bin $f_j$ with $f_i$ with probability $P_0$ if $C_{ij}\geq 1-\delta_0$. Here, $1-\delta_0$ is a design parameter indicating the minimum correlation among the channel vectors in a cluster. Then, the threshold on the maximum value of $\hat{D}^{ij}_c$ is $\gamma^{j}_0=\delta_0\sigma_j Q^{-1}(P_0)+\delta_0\mu_{j}$, where $Q^{-1}(.)$ is the inverse Q-function. Therefore, frequency bin $f_j$ is clustered with $f_i$ if 
$\hat{D}_c^{ij}= \text{Tr}\left((\mathbf{\hat{U}}^{f_i})^H \left(\mathbf{\hat{R}}^{f_j} -\mathbf{R}^{f_j}_w \right)\mathbf{\hat{U}}^{f_i}\right) \leq \gamma^j_0$.
The estimates of the mean and the variance are obtained using the non-asymptotic estimate of the covariance as $\hat{\mu}_{j} = \text{Tr}(\mathbf{\hat{R}}^{f_j} - \mathbf{R}^{f_j}_w)$ and $\hat{\sigma}^2_{j} = (\hat{\mu}_{j}^2 + \sigma^4_w)/T$.

\subsection{Clustering algorithm}
\label{section:clustering_algorithm}

We assume that the BS knows the set of frequency bins $\mathcal{F}_a$ where OUs are active. In the clustering algorithm, the set of clusters is initialized as an empty set: $\mathcal{S} = \phi$. Let $\mathcal{A}$ be the set of frequency bins which are not clustered. It is initialized as $\mathcal{A} = \mathcal{F}_a$. Let $\mathcal{B}_1,...,\mathcal{B}_n$ denote the contiguous occupied frequency bands in $\mathcal{A} = \{\mathcal{B}_1,...,\mathcal{B}_n\}$. The clustering algorithm first computes the number of signals $\hat{L}_{f_1}$ and the null space $\mathbf{\hat{U}}^{f_1} \in \mathbb{C}^{M \times (M-\hat{L}_{f_1})}$ at the center frequency $f_1=\lfloor \sum_i{\mathcal{B}_1(i)} / |\mathcal{B}_1| \rfloor$ of the first contiguous band $\mathcal{B}_1$. A frequency cluster of correlated bins $\mathcal{S}'$ is formed with center bin $f_1$ at the center of the cluster.
The algorithm then iteratively checks whether the adjacent frequency bins $f_j = f_1  \pm \Delta f$ can be clustered with $f_1$ using the statistic $\hat{D}^{1j}_c$. Once the cluster $\mathcal{S}'$ is formed around the bin $f_1$, a common null space matrix $\mathbf{U}^{\mathcal{S}'} \in \mathbb{C}^{M \times (M-\hat{L}_{f_i})}$ is computed for the cluster by computing the EVD of the normalized sum $\mathbf{R}^{\mathcal{S}'}$ of covariance matrices in the cluster: $\mathbf{R}^{\mathcal{S}'} = \sum_{f \in \mathcal{S}'} \frac{1}{\hat{\mu}_{j}}(\mathbf{\hat{R}}^{f} -\mathbf{{R}}^{f}_w)$. The covariance matrices are normalized by the average power in order to provide equal weight for covariance matrices in the cluster irrespective of the power received in the bin. This is due to the fact that the null space of the channel vectors $\mathbf{h}^f_l$ is independent of the received power $p^f_l ||\mathbf{h}^f_l||^2$ in that bin. Frequency bins clustered in $\mathcal{S}'$ are then removed from $\mathcal{A}$ and the algorithm proceeds by rearranging contiguous bands in $\mathcal{A}$, as shown in Algorithm \ref{alg:clustering_algorithm}. Note that there are two EVD computations for each cluster with more than one frequency bin in steps 4 and 10 of the algorithm. While for clusters with only one frequency bin, there is only one EVD computation, as step 10 becomes redundant. Therefore, total number of EVD computations is $|\mathcal{S}|+ |\mathcal{S}_1|$, where $\mathcal{S}_1$ is the set of clusters with only one frequency bin, while the number of the number of matrices $\mathbf{U}^{\mathcal{S}'}$ computed is equal to number of clusters $|\mathcal{S}|$.

{\scriptsize
	\begin{algorithm}[ht]
		\caption{Clustering algorithm}
		\label{alg:clustering_algorithm}
		\begin{algorithmic}[1]
			\State Initialization: $\mathcal{S}=\phi$, $\mathcal{A} = \mathcal{F}_a = \{\mathcal{B}_1,...,\mathcal{B}_n\}$.
			\While{$\mathcal{A} \neq \phi$}
			\State Compute $f_1 = \lfloor \sum_i{\mathcal{B}_1(i)} / |\mathcal{B}_1| \rfloor$.
			\State Compute $\hat{L}_{{f}_1}$, $\mathbf{\hat{U}}^{f_1}$ using EVD of $\mathbf{\hat{R}}^{f_1} - \mathbf{R}^{f_1}_w$.
			\Statex Form cluster $\mathcal{S}'$ around bin $f_1$:
			\State Initialize: $\mathcal{S}'=\phi$, $\Delta f =1$. Set $f_j=f_1 +\Delta f$. 
			\While{$\hat{D}^{1j} \leq \gamma^{j}_0$}
			\State $\mathcal{S}' \leftarrow \mathcal{S}' \cup \{f_j\}$.
			\State Increment: $f_j = f_j+\Delta f$.
			\EndWhile
			\Statex~~ (Repeat steps 6 to 9 with $\Delta f=-1$.)
			\State Compute $\mathbf{U}^{\mathcal{S}'}\in \mathbb{C}^{M\times M-\hat{L}_{f_1}}$ for cluster $\mathcal{S}'$ using EVD of $\mathbf{R}^{\mathcal{S}'} = \sum_{f \in \mathcal{S}'} \frac{1}{\hat{\mu}_{j}}(\mathbf{\hat{R}}^{f} -\mathbf{{R}}^{f}_w)$.
			\State Add $\mathcal{S}'$ in set of clusters: $\mathcal{S} \leftarrow \mathcal{S} \cup \mathcal{S}'$.
			\State Exclude the clustered bins from $\mathcal{A}$: $\mathcal{A} \leftarrow \mathcal{A} \backslash \mathcal{S}'$.
			\State Rearrange contiguous bands in $\mathcal{A} =\{\mathcal{B}_1,...,\mathcal{B}_n\}$.
			\EndWhile
		\end{algorithmic}
\end{algorithm}}
\vspace*{-4mm}

\section{Simulation results}
\label{sec:simulation}
\begin{figure*}
	\begin{subfigure}[b]{0.5\textwidth}
			\includegraphics[width=1.2\columnwidth]{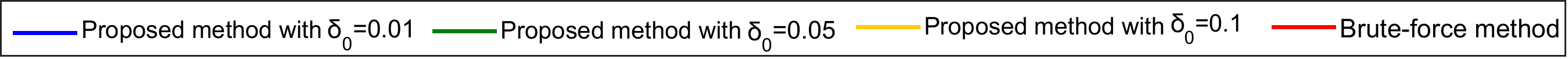}	
			\label{fig:legend}
			\vspace{-5mm}
		\end{subfigure}		
	\\\vspace{-3.5mm}\\\centering
	\begin{subfigure}[b]{0.3\textwidth}
		\includegraphics[width=1.1\columnwidth]{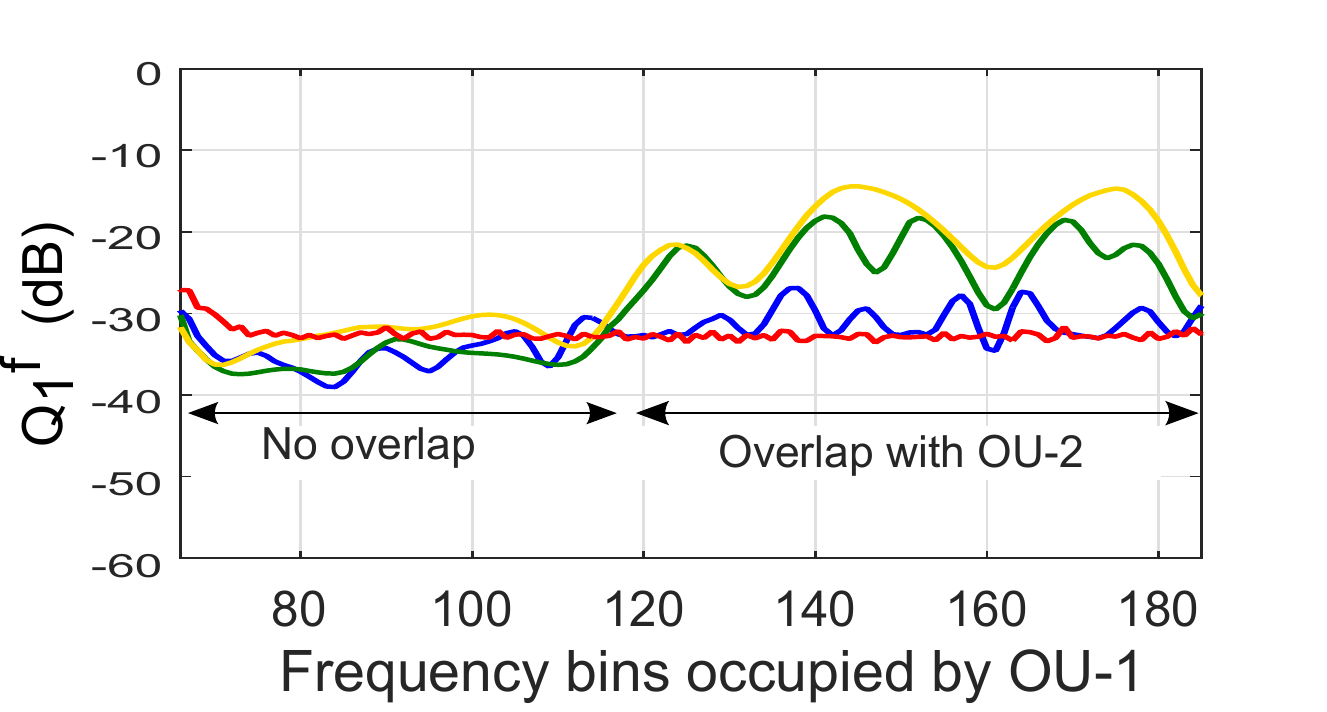}
		\caption{{\footnotesize Null to OU-1 in urban microcell.}}
		\label{fig:pu_1_ch_13}
		\vspace{-1mm}
	\end{subfigure}
	\begin{subfigure}[b]{0.28\textwidth}
		\includegraphics[width=1.1\columnwidth]{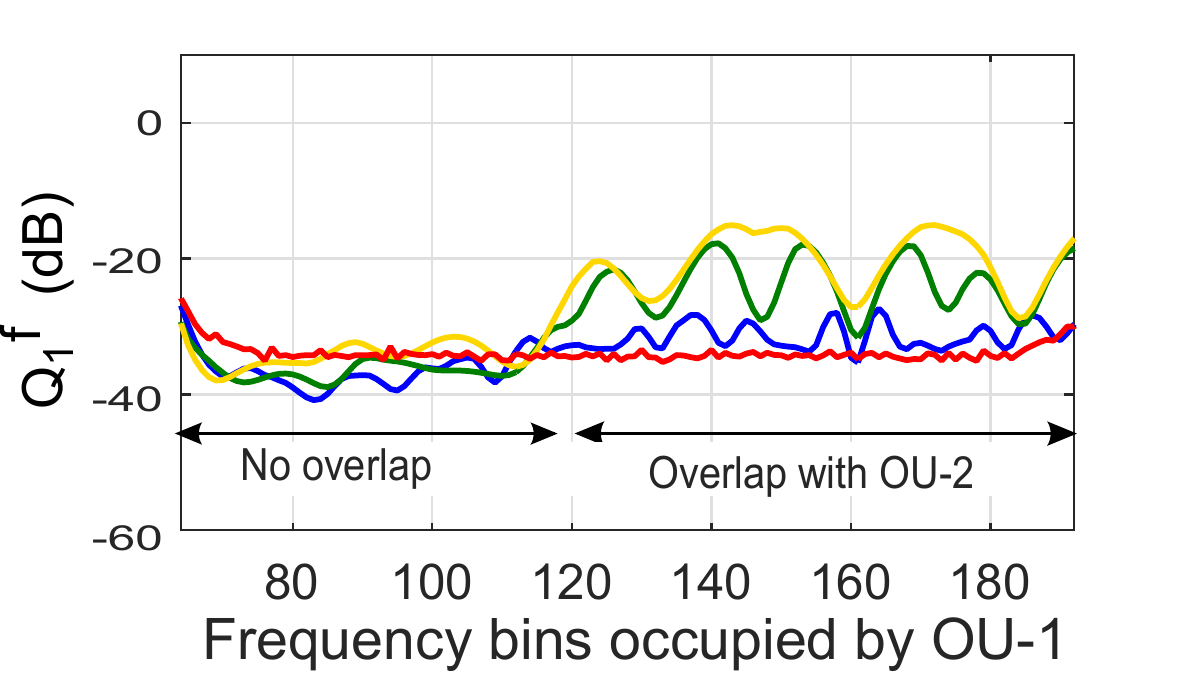}
		\caption{{\footnotesize Null to OU-1 in EPA model.}}
		\label{fig:pu_1_ch_5}
		\vspace{-1mm}
	\end{subfigure}
	\begin{subfigure}[b]{0.28\textwidth}
		\includegraphics[width=1.09\columnwidth]{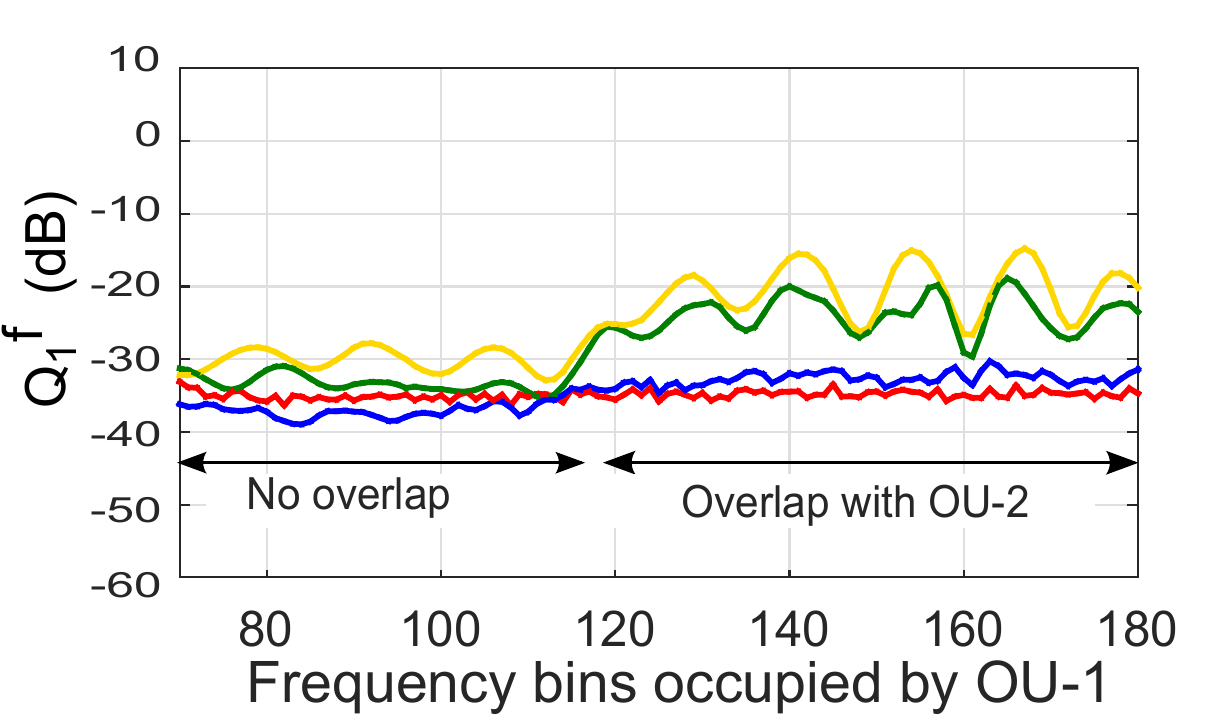}
		\caption{{\footnotesize Null to OU-1 in ETSI B model.}}
		\label{fig:pu_1_ch_9}
		\vspace{-1mm}
	\end{subfigure}
	\\
	\begin{subfigure}[b]{0.3\textwidth}
		\includegraphics[width=1.1\columnwidth]{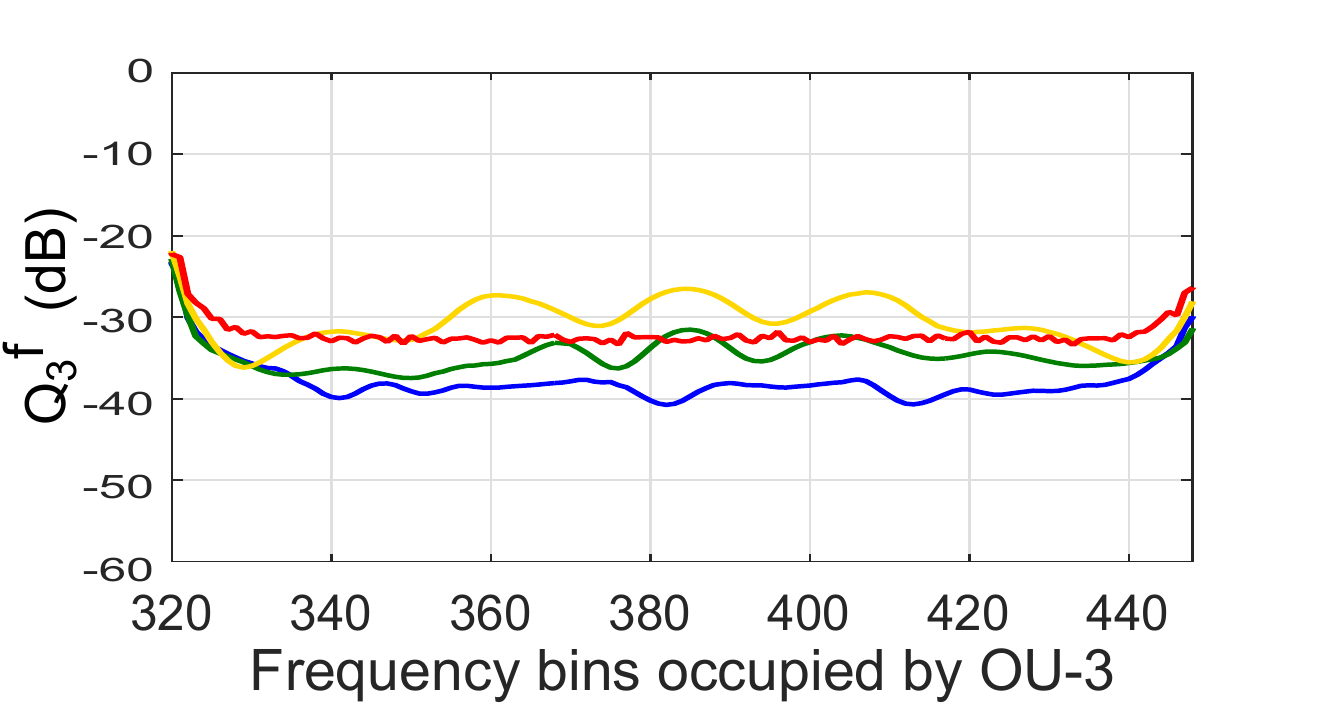}
		\caption{{\footnotesize Null to OU-3 in urban microcell.}}
		\label{fig:pu_3_ch_13}
		\vspace{-1mm}
	\end{subfigure}
	\begin{subfigure}[b]{0.27\textwidth}
		\includegraphics[width=1.1\columnwidth]{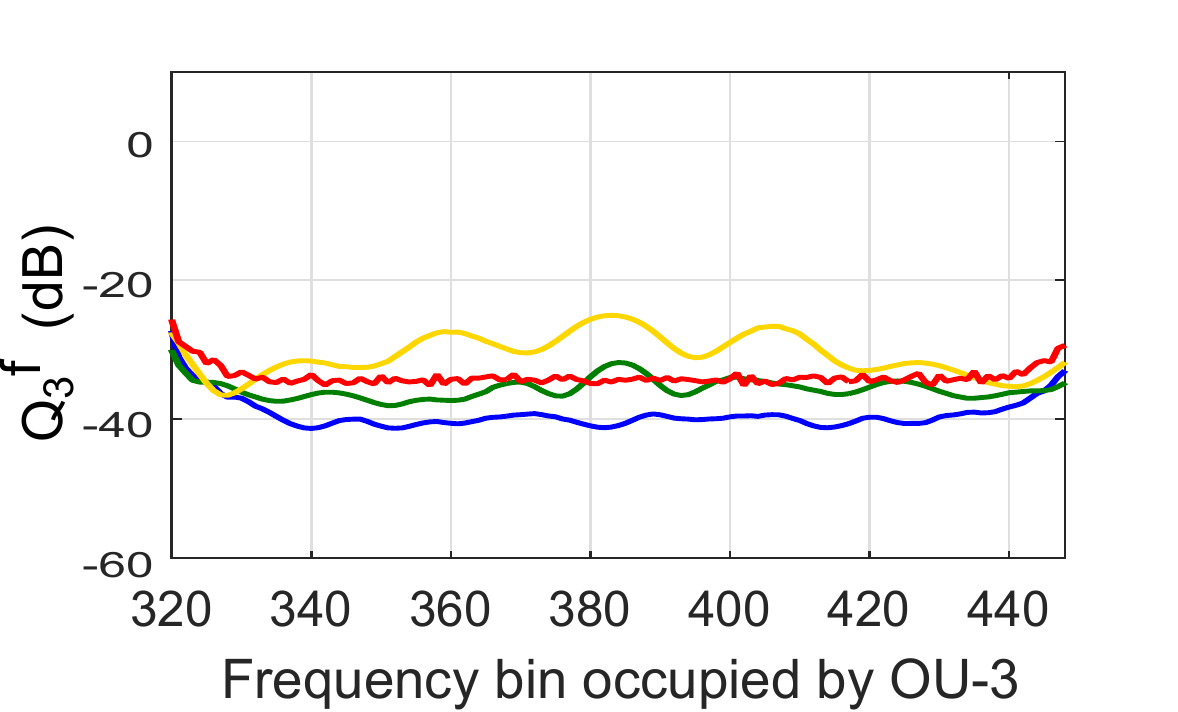}
		\caption{{\footnotesize Null to OU-3 in EPA model.}}
		\label{fig:pu_3_ch_5}
		\vspace{-1mm}
	\end{subfigure}
	\begin{subfigure}[b]{0.27\textwidth}
		\includegraphics[width=1.1\columnwidth]{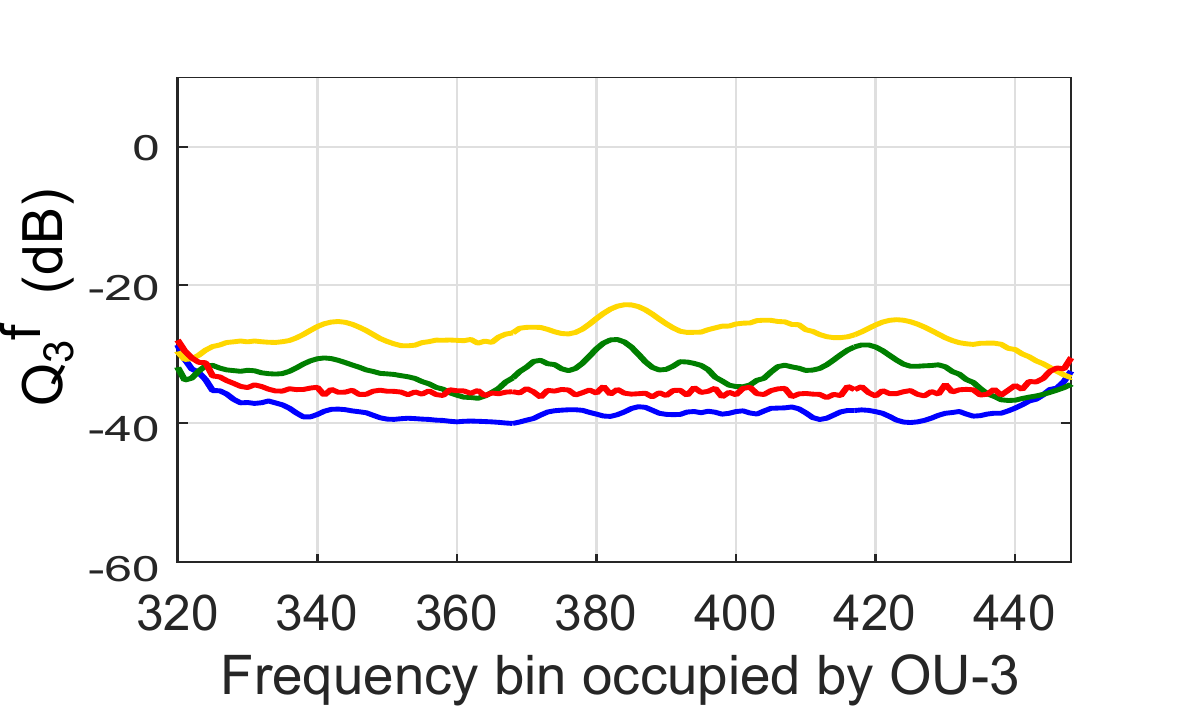}
		\caption{{\footnotesize Null to OU-3 in ETSI B model.}}
		\label{fig:pu_3_ch_9}
		\vspace{-1mm}
	\end{subfigure}	
	\caption{\footnotesize Relative quality of nulls $Q^f_l=20\log_{10}\left({||(\mathbf{U}^{\mathcal{S}'})^H \mathbf{h}^f_l||}/{||\mathbf{h}^f_l||}\right)$ toward OU-1 and OU-3 under three channel models.}
	\label{fig:null_quality}
	\vspace{-4mm}
\end{figure*}

\begin{figure}
	\centering
	\includegraphics[width=\columnwidth]{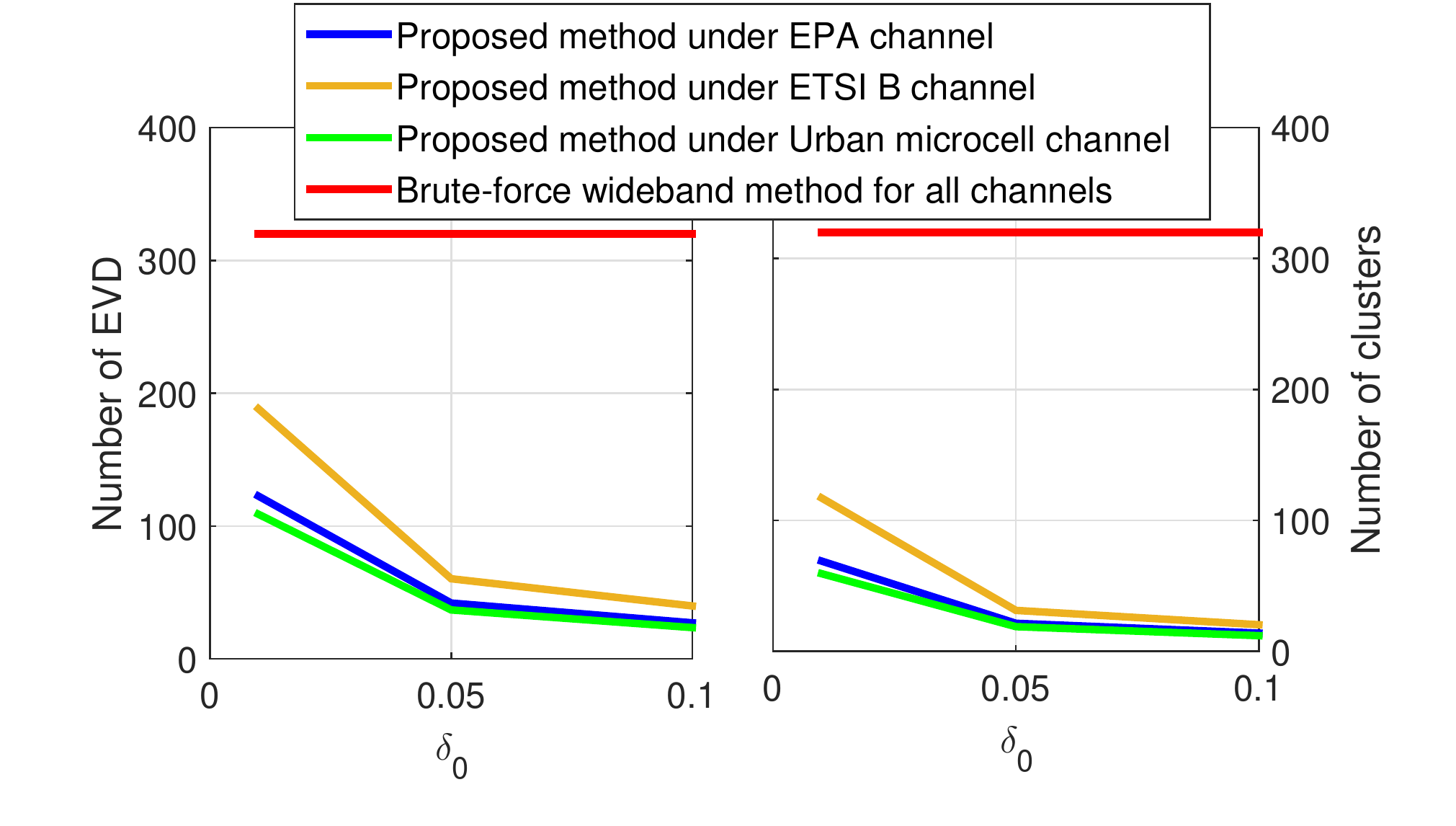}
	\vspace{-9mm}
	\caption{{\footnotesize Number of clusters and EVD computations }}				
	\vspace{-8mm}
	\label{fig:number_clusters_evd}
\end{figure}

We evaluate the performance of the algorithm under three channel models, namely urban microcell, extended pedestrian A (EPA), and ETSI B with RMS delay spreads 36.17ns, 43.13ns, and 98.99ns, respectively. The BS observes baseband frequencies from $-10$MHz to $10$MHz at sampling rate $1/T_s=20$MHz. The number of FFT points are $F=512$, and noise variance is $\sigma^2_{w}=1$. The probability of clustering is set to $P_0=0.95$, while the non-asymptotic covariance matrix is estimated using $T=100$ samples. The number of antennas at the BS is $M=8$. We consider that the BS receives signals from $L=3$ OUs with baseband center frequencies $-5$MHz, $-2.5$MHz and $5$MHz, respectively. The OUs transmit OFDM signals with bandwidths $5$MHz. Assuming 256-th bin corresponds to the center frequency, the signals transmitted from OU-1, OU-2, and OU-3 occupy frequency bins in the range 64 to 192, 128 to 256, and 320 to 448, respectively. Note that the signals received from OU-1 and OU-2 have 50\% overlap while the signal received from OU-3 has no overlap with any other signal. The transmit powers are scaled such that the received SNRs are exponential random variables with means $10$dB over 10000 channel realizations. We compare the relative quality of null $Q^f_l$ generated by the proposed algorithm with brute-force method where the null space matrices $\mathbf{U}^f$ are computed separately at each active bin $f\in \mathcal{F}_a$. This brute-force approach is an extension of the methods proposed in \cite{gao2010a} and \cite{tsinos2013a} to a wideband scenario. Note that lower $Q^f_l$ indicates better quality of null. The quality of the null for the OU-1 and OU-3 under the three channel models is shown in Fig. \ref{fig:null_quality}. The results for OU-2 are not shown as they resemble OU-1 results due to similar spectral overlap. Note that the value of $Q^f_l$ depends on the number of samples $T$ used to compute the non-asymptotic covariance matrix. In this work, we are interested in comparing $Q^f_l$ for the proposed and the brute-force method for same $T$. It can be observed that the quality of nulls generated using the proposed method degrades as the $\delta_0$ increases. This is because larger $\delta_0$ results in larger, but fewer  clusters, as shown in Fig. \ref{fig:number_clusters_evd}, which means that the common null space $\mathbf{U}^{\mathcal{S}'}$ does not represent null space in all frequency bins in the cluster. Further, the quality of null for $\delta_0=0.1$ is poor for OU-1 in the frequency bins above 110 where the signal received from OU-2 starts overlapping with OU-1 signal. This means that larger cluster does not produce sharp nulls in the frequency bins with spectrally overlapping signals. In such case, $\delta_0=0.01$ should be used for clustering. 

The main advantage of the proposed clustering algorithm can be seen in Figs. \ref{fig:pu_3_ch_13}, \ref{fig:pu_3_ch_5}, and \ref{fig:pu_3_ch_9}, where the quality of nulls is shown for OU-3 in frequency bins 320 to 448. Note that there is no spectral overlap with any other signal in this band. It can be observed that the matrices $\mathbf{U}^{\mathcal{S}'}$ generated with $\delta_0=\{0.01,0.05\}$ provide nulls with lower $Q^f_l$ as compared with brute-force wideband method. We can also see in Fig. \ref{fig:number_clusters_evd} that the number of EVD computations is reduced to 1/2 and 1/4 for $\delta_0=0.01$ and $\delta_0=0.05$, respectively, as compared to the brute-force method. As mentioned before, the number of matrices $\mathbf{U}^{\mathcal{S}'}$ required to be computed is equal to the number of clusters and it reduces to less than 1/3 as compared to the existing method as shown in Fig. \ref{fig:number_clusters_evd}. Finally, the number of clusters and the number EVD computations reduce as the RMS delay spread of the channel reduces. This is because smaller delay spread results in higher correlations among channel vectors in adjacent frequency bins for same value of $T_s$ and $F$, as shown in Fig. \ref{fig:channel_correlation}, which in turn results in fewer clusters for same value of $\delta_0$.

\vspace{-2mm}
\section{Conclusion}
\label{sec:conclusion}
We proposed an unsupervised frequency clustering algorithm for null space computation for wideband channels between a BS and users associated with other coexisting network in the same frequency band. The proposed algorithm clusters frequency bins with correlated channels using the received signal on the BS antenna array without any prior knowledge of the channels or training signals. A common null space matrix is computed for clustered frequency bins to reduce the computational complexity. The results show that the proposed algorithm has significantly lower computational complexity if the RMS delay spread of the channel is smaller for same sampling duration and the number of FFT bins.


\bibliographystyle{IEEEbib}
\bibliography{GlobalSIP2017_references}

\end{document}